\documentclass[twocolumn,a4paper]{article}
\usepackage[style=numeric,sorting=none,backend=bibtex]{biblatex} 
\bibliography{Paper}
\usepackage{booktabs}
\usepackage{graphicx}
\usepackage{abstract}
\usepackage{hyperref}
\hypersetup{colorlinks=false,
pdftitle={Wayside crosses -- a case study.},
citebordercolor={1 1 1},
linkbordercolor={1 1 1},
urlbordercolor={1 1 1}}

\begin{document}

\title{Data coverage, richness, and quality of OpenStreetMap for special interest tags: wayside crosses -- a case study.}

\author{Philipp Weigell\thanks{arxiv@weigell.com}}

\twocolumn[
  \begin{@twocolumnfalse}
    \maketitle
    \begin{abstract}
Volunteered Geographic Information projects like OpenStreetMap which allow accessing and using the raw data, are a treasure trove for investigations – e.\,g. cultural topics, urban planning, or accessibility of services. Among the concerns are the reliability and accurateness of the data. While it was found that for mainstream topics, like roads or museums, the data completeness and accuracy is very high, especially in the western world, this is not clear for niche topics. Furthermore, many of the analyses are almost one decade old in which the OpenStreetMap-database grew to over nine billion elements.

Based on OpenStreetMap-data of wayside crosses and other cross-like objects regional cultural differences and prevalence of the types within Europe, Germany and Bavaria are investigated. For Bavaria, internally and by comparing to an official dataset and other proxies the data completeness, logical consistency, positional, temporal, and thematic accuracy is assessed. Subsequently, the usability for the specific case and to generalize for the use of OpenStreetMap data for niche topics. 

It is estimated that about one sixth to one third of the crosses located within Bavaria are recorded in the database and positional accuracy is better than 50\,metres in most cases. In addition, linguistic features of the inscriptions, the usage of building materials, dates of erection and other details deducible from the dataset are discussed. It is found that data quality and coverage for niche topics exceeds expectations but varies strongly by region and should not be trusted without thorough dissection of the dataset.\\\\
    \end{abstract}
  \end{@twocolumnfalse}
  ]
\saythanks 

\section{Introduction}
Volunteered Geographic Information (VGI) projects \cite{Goodchild2007} like OpenStreetMap \cite{OpenStreetMap}, which allow accessing and using the raw data under the Open Data Commons Open Database License \cite{odbl}, offer a treasure trove of data to investigate different topics. This includes, among others, cultural topics, urban planning tasks, or accessibility of services. Among the concerns are the reliability and accurateness of the data \cite{Antoniou_2015,Neis_2013,Jackson_2013,Mooney2010,haklay2010}. While it was found that for mainstream topics, like roads \cite{Barrington2017} or museums \cite{Balducci2021} the data completeness and accuracy is very high, especially in the western world, this is not clear for niche topics. Also many of the studies are one decade old in which the database grew significantly to over nine billion mapped elements \cite{osm_wiki_stats}. 

By design and its guidelines \cite{osm_wiki_contribution} OpenStreetMap is open for any geographical information, which is verifiable on the ground. This reaches from roads, to the colour of window frames in the third-floor of a building. However, as the data is contributed voluntarily and uncoordinated, especially niche topics will have a very inhomogeneous coverage. In addition, inherent biases have to be kept in mind. E.\,g. \cite{Gardner2019} found that information contributed by persons identifying as men is different from that contributed by persons identifying as women. The same can be expected by age, cultural background, education and furthermore. 

All information is recorded in key=value pairs which are called tags on three types of objects: nodes, ways and relations. While nodes denote point-like objects, ways, which are composed of several nodes, describe extended objects like streets or rivers. If ways are closed they can also describe buildings or other two-dimensional objects. Finally, relations, which are composed out of ways and/or nodes, describe more complex objects, like for example a country boundary including enclaves. The object type or further details are organized in the tags, e.\,g. keys like \textsc{material} and values like \textsc{wood} \cite{osm_wiki_elements}. As each object may have several tags, information can be faceted and enable complex studies. 

This study tries to understand for a niche topic, i.\,e. wayside crosses and other cross-like objects, how rich the information in the database actually is, how much it varies with regional or other factors and how far one can rely on it \cite{ThebaultSpieker2018}. The main focus will be on the state of Bavaria in Germany, but first the global, the European, and the German situation will be discussed. 

All data discussed was accessed on May 23$^{\mathrm{rd}}$ 2023 via the overpass turbo API \cite{Raifer2013}.
\section{Global situation}
On a global scale 170311 objects are mapped (i.\,e. stored in the database) as \textit{historic=wayide\_cross} (\textsc{hwc}). \cite{osm_wiki_hwc}  The vast majority, 169511, of these are nodes. It can be assumed that most of the other objects, which are mapped more complex, i.e. as way or relation, are mapped incorrectly or are indeed other kinds of place to worship as for example a wayside shrine. This is also supported by the official documentation, which encourages only the use of nodes. As the non-adherence to the \textsl{correct} tagging is only 0.5\,\%, this study will concentrate on nodes to ensure consistent data. 

In addition, there are five other node-types available in the OpenStreetMap dataset which are related. First, \textit{historic=wayside\_shrine} (\textsc{hws}, 103746 objects) \cite{osm_wiki_hws}. Especially in the Western context these often refer to so called \textit{Bildstöcke}, which depending on the strictness of definition can be counted as wayside cross as they serve a very similar purpose. 

\begin{figure}
\centering
\includegraphics[width=1.00\columnwidth]{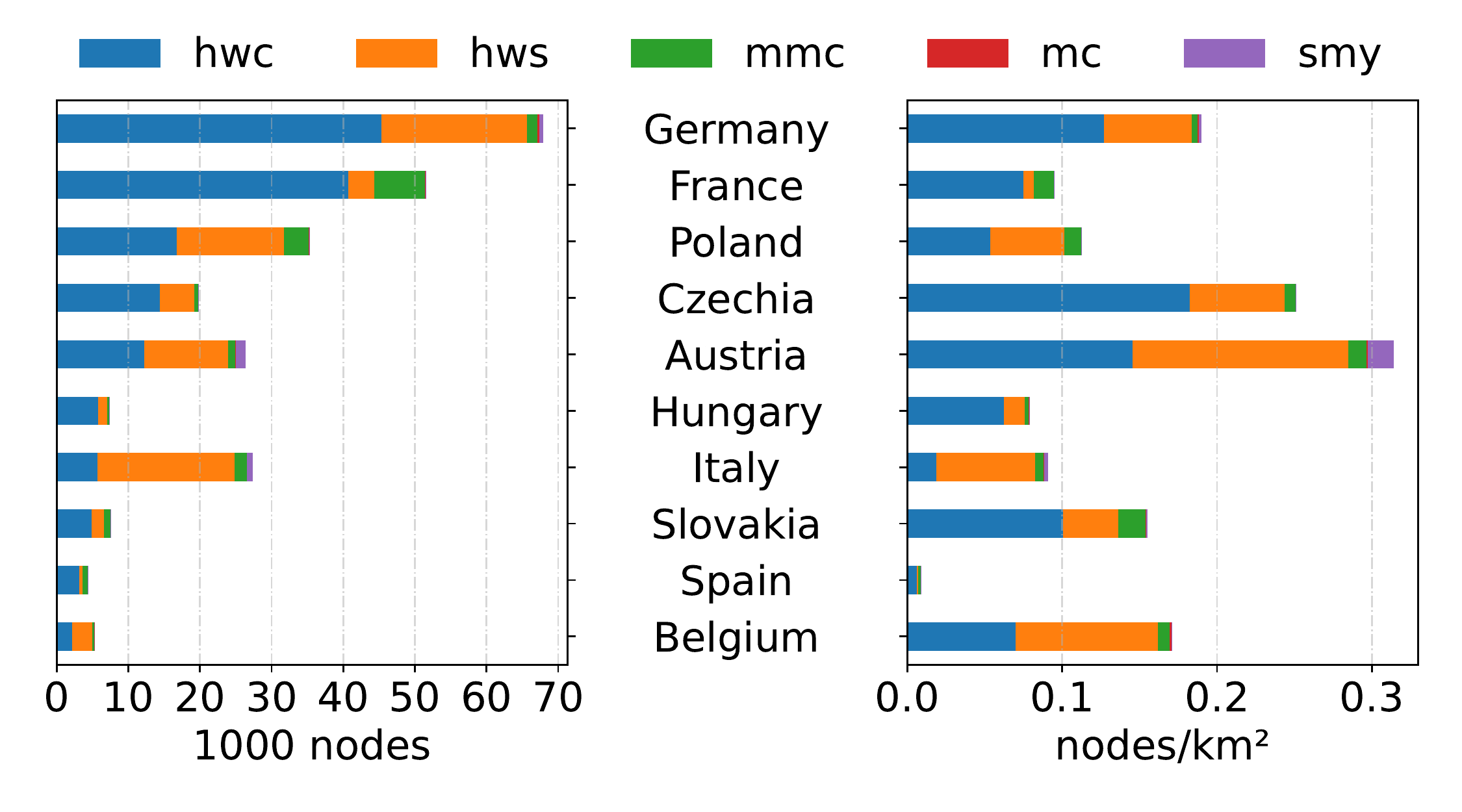}
\caption{Absolute count (left) and per area count (right) of \textsc{hwc}, \textsc{hws}, \textsc{mmc}, \textsc{mc}, \textsc{mtc}, and \textsc{smy} nodes for the top-10 countries with the most \textsc{hwc} nodes.}
\end{figure}

Second, there is the tag for crosses without historical value as seen by the person registering it to the database: \textit{man\_made=cross} (\textsc{mmc}, 21985 objects) \cite{osm_wiki_mmc} and third summit crosses denoted by summit:cross=yes (\textsc{smy}, 3285 objects) \cite{osm_wiki_smy}. Last, there is memorial=cross (\textsc{mc}, 1075 objects) and memorial:type=cross (\textsc{mtc}, 59 objects). Both denote memorials for special events or persons with the latter even lacking a formal definition in the OpenStreetMap-Wiki. \cite{osm_wiki_mc} As especially for the latter three object types the folksonomy seems not yet to be well defined, they tend to be combined in a few cases. Table \ref{combinations} summarizes how often each tag is combined on node-objects with each other tag, normalized to the number of occurrences of the tag in question. As can be seen, when analysing \textsc{hwc}, contamination is only half a percent, and \textsc{hwc} and \textsc{hws} are clearly distinct. As the focus will be on \textsc{hwc} in the following this contamination will be treated as negligible and thus disregarded. 

As pointed out in \cite{Mocnik2017} it is expected, that over time the specificity will increase.  However, over a decade after \cite{Mooney_2012} one of the main problems found there, still remains: data is not consistently tagged in the OpenStreetMap-database. This of course is also rooted in the fact, that a typical contributor will in many cases not be able to distinguish if a cross has a historic value or is "just" a memorial for example. Here the community, i.e. the active contributors, will need to find a less complex but more exact definition if tagging shall converge at one point.  

\begin{table}
\small\sf\centering
\caption{Percentage of objects bearing the other key in relation to the total number objects with that object type.} 
\label{combinations}
\begin{tabular}{lrrrrrr}
\toprule
{} &     \textsc{hwc} &     \textsc{hws} &     \textsc{mmc} &      \textsc{mc} &     \textsc{mtc} &     \textsc{smy} \\
\midrule
\textsc{hwc} & - &   0.0\% &   0.5\% &   0.0\% &   0.0\% &   0.0\% \\
\textsc{hws} &   0.0\% & - &   0.0\% &   0.0\% &   0.0\% &   0.0\% \\
\textsc{mmc} &   3.6\% &   0.2\% & - &   0.2\% &   0.0\% &   2.0\% \\
\textsc{mc}  &   2.4\% &   0.0\% &   3.2\% & - &   0.7\% &   0.4\% \\
\textsc{mtc} &   3.4\% &   0.0\% &   1.7\% &  13.6\% & - &   0.0\% \\
\textsc{smy} &   2.2\% &   0.0\% &  13.1\% &   0.1\% &   0.0\% & - \\
\bottomrule
\end{tabular}
\end{table}

\section{European situation}
The top ten countries in terms of mapped \textsc{hwc}-nodes are all European, with in absolute terms the most nodes in Germany, France, and Poland. This is both to be expected as, first, Europe has a strong Christian tradition. Second, the data-completeness is high if roads are employed as a proxy \cite{Barrington2017} and third, the top three countries are among the larger ones in terms of area within Europe. A notable miss here is the predominantly protestant UK. 

If additionally, wayside shrines are considered, which are assumed to be predominantly Christian in European countries, Austria and Italy are significantly closer to Poland. This reflects different cultural traditions which evolved in the medieval culture of Europe, like the preference for wayside shrines in Southern Tyrol \cite{Timmermann2012,KING_1985}.  Although, many crosses are erected for mourning, e.\,g. after car accidents, with less focus on the religious aspect \cite{Aka2007,Przybylska2021,Klaassens_2013,Ne_porov__2013}, \textsc{hwc}-nodes are only found in France and all other tags are negligible. 

Normalized to the area, the order changes. While Germany is still in the top-3 France drastically falls behind, even when disregarding the overseas territories. This could be rooted in the secular tradition in France \cite{WILLAIME_1998} and/or incomplete data. As similarly the strongly Catholic Poland with a strong tradition in erecting crosses \cite{Przybylska2016} and the likewise Catholic Spain, fall behind, the latter seems more likely or at least a strong factor. For a definite answer larger field studies would be needed. Smaller countries, with a strong Christian tradition like Czechia and Austria but also Belgium exhibit high densities, especially if we consider wayside shrines as well. Specifically, Belgium is noteworthy here, as its neighbour country The Netherlands have much lower numbers. As both countries can be assumed to have a similar data coverage \cite{Barrington2017} this probably stems from the fact that The Netherlands have a protestant tradition and that the cross symbol has significantly different importance, when compared to Belgium \cite{MacConville_2010,Przybylska2020,Klaassens_2013}.

Still, while in all these European countries more than half of the population identifies as Christians \cite{CIA2020}, no clear correlation with the percentage of Christians nor Catholics, can be identified. While this to some extend reflects different cultural rites on national and regional scales, it also indicates missing data, as we show for Germany, an area with better coverage, the correlation is strong (cf. Figure \ref{fig:germrel}). 

In reverse, the concentration of the data in Europe, though expected given the Christian background, reflects the incompleteness of the OpenStreetMap-data in less wealthy regions of the world. As projects with the aim of improving map data in less wealthy regions, like Missing Maps by the red cross and red crescent organizations, focus on "mainstream topics" \cite{Scholz_2018}, this imbalance most likely will persist, till vital local communities emerge.

\section{German situation}
\begin{figure}
\centering
\includegraphics[width=1.00\columnwidth]{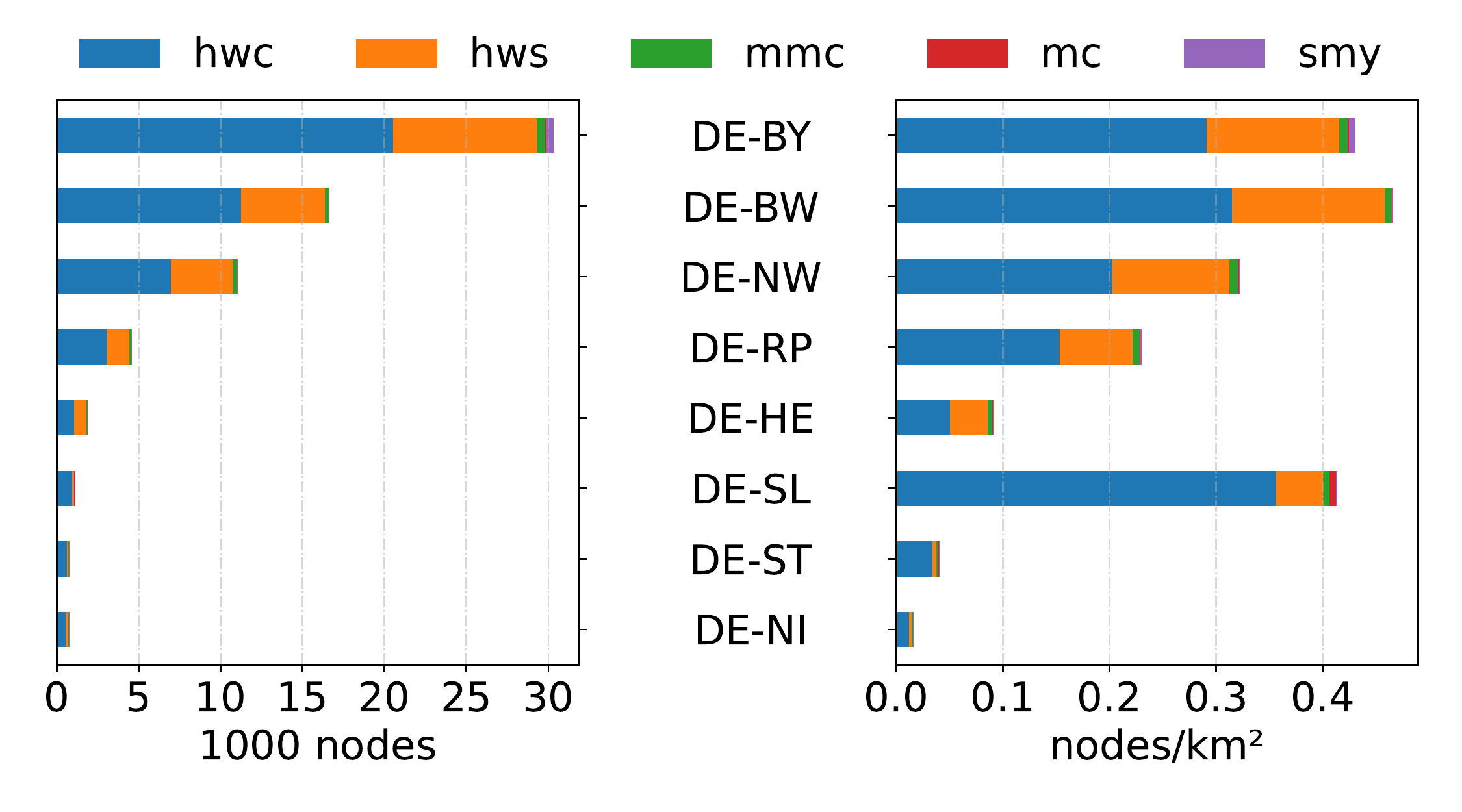}
\caption{Absolute count (left) and per area count (right) of \textsc{hwc}, \textsc{hws}, \textsc{mmc}, \textsc{mc}, \textsc{mtc}, and \textsc{smy} nodes for German states with at least 50 mapped \textsc{hwc} nodes. State names are abbreviated by their ISO-3166-code.}
\label{distribution_germany}
\end{figure}
\begin{figure}
\centering
\includegraphics[width=1.00\columnwidth]{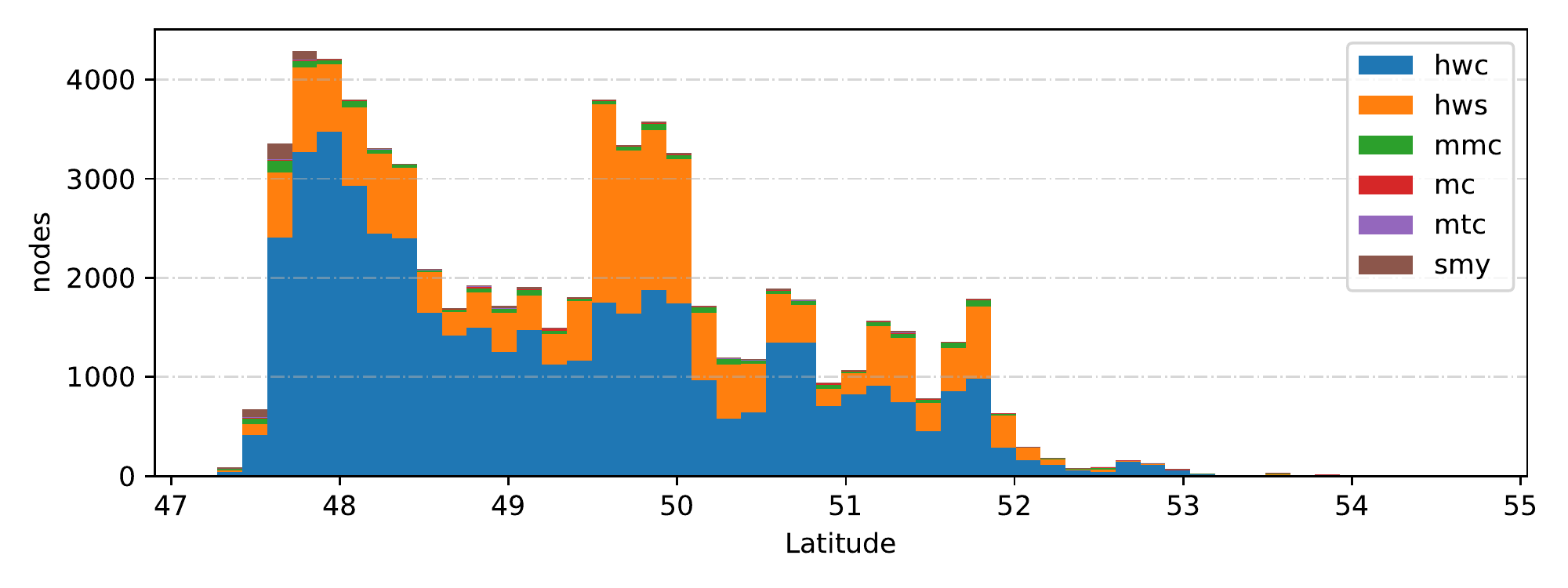}
\caption{Count of \textsc{hwc}, \textsc{hws}, \textsc{mmc}, \textsc{mc}, \textsc{mtc}, and \textsc{smy} nodes by Latitude within Germany.}
\label{northsouthgermany}
\end{figure}

Next, we will focus on Germany, where “traditional” wayside crosses are predominant and more than one quarter of the wayside cross-objects mapped worldwide are located. Within Germany almost half of the objects are concentrated in Bavaria (BY), followed by the second southern state of Baden-Württemberg (BW), see Figure \ref{distribution_germany} (left). North Rhine-Westphalia (NW) is the last state which has more than 5000 crosses recorded. This aggregation in the south is also reflected in a strong correlation of the number of \textsc{hwc}-nodes with increasing Latitude (see Figure \ref{northsouthgermany}). Roughly at the start of the North German Plain almost no nodes occur in the database. As the data quality within Germany can be assumed to be uncorrelated with the Latitude \cite{Ludwig2011} this hints to a correlation with the the predominant denominations in these regions.
In combination with \textsc{hws} a second maximum is around 50$^\circ$ North.  We will discuss this further later when focusing on Bavaria. 

Similar to the global and European scale, including the other four types, \textsc{mmc}, \textsc{mc}, \textsc{mtc}, and \textsc{smy}, does not change the picture as their admixture is low in Germany; pointing to a better tag-definition or a stricter folksonomy within the German OpenStreetMap community. In addition to the decrease from south to north, the former division of Germany into West and East, with the secular doctrine in the Eastern part seems to be reflected in the data as no state of the former German Democratic Republic (GDR) reaches 1000 nodes, and four states not even cross the bar of 50 nodes. 

Normalizing for the area (see Figure \ref{distribution_germany} (right)), Saarland (SL), a region with a particular strong French history and influence but also with the highest rate of Catholics among its population \cite{SAeBL2014}, yields the highest density. This is remarkable, as we have seen before for France the contrary is true. If this is rooted in different focus of the respective OpenStreetMap communities, the different rate of Catholics or in other different cultural rites can only be investigated by integrating additional, external data.

Furthermore, the two southern states Bavaria and Baden-Württemberg exhibit almost the same densities of nodes for the tags investigated in this paper. 

\begin{figure}
	\centering
		\includegraphics[width=1.00\columnwidth]{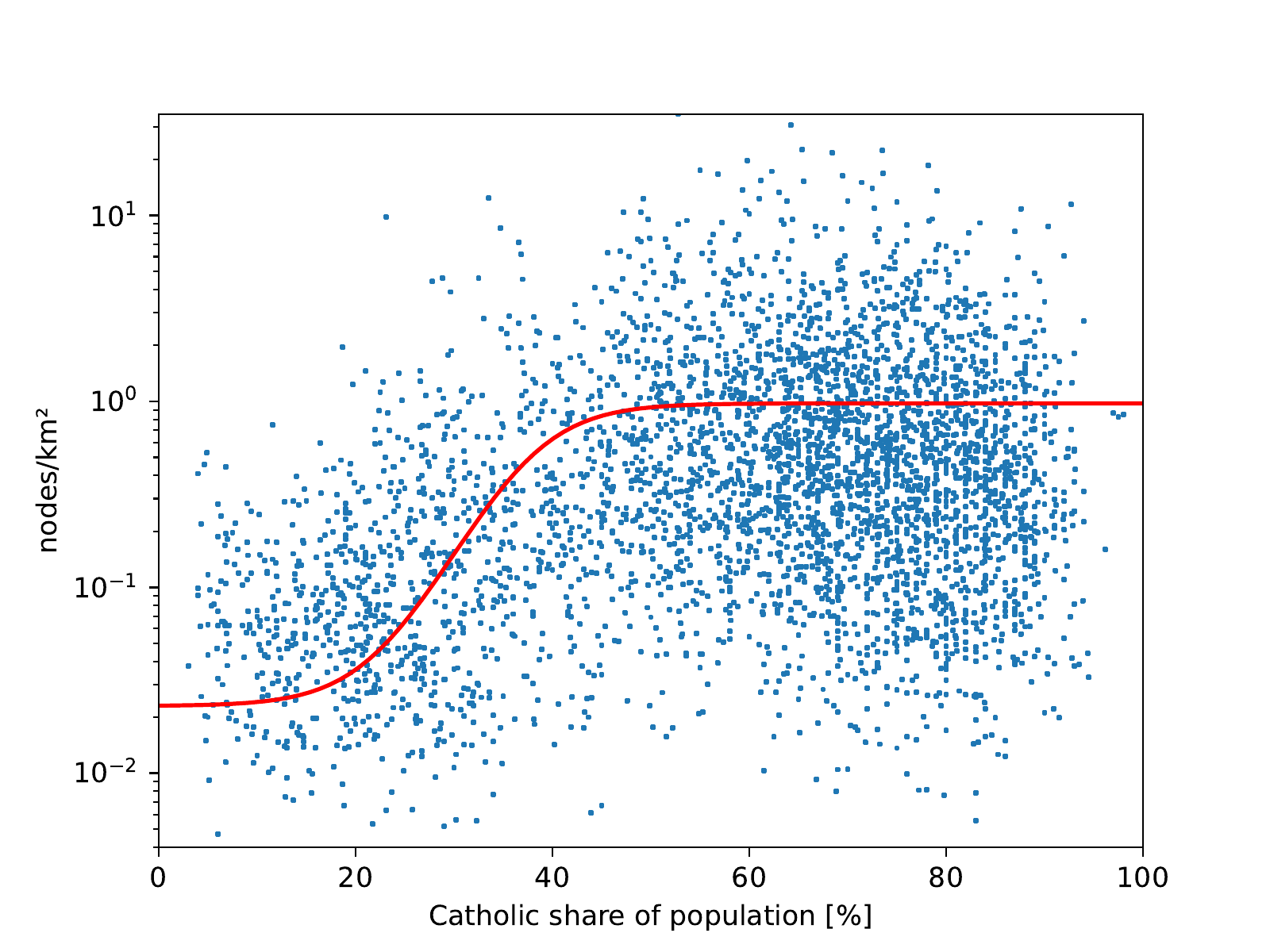}
	\caption{Areal densities for the sum of all \textsc{hwc}, \textsc{hws}, \textsc{mmc}, \textsc{mc}, \textsc{mtc}, and \textsc{smy} nodes per county in Germany versus the share of Catholics within the counties population. Additionally, the least square fit of a logistic function ($f(x) = a / (1 + \textrm{e}^{(-k*(x-x^0))})+b$)
	is drawn as red line. Note the logarothmic scaling on the ordinate.}
	\label{fig:germrel}
\end{figure}

To check if the areal densities of mapped \textsc{hwc} and/or the sum of all studied objects depend on the rate of Christians and/or Catholics, these were determined for all German counties and a logistic function was fitted. As measure for the rate of Catholics and Protestant the official percentage of the most recent general census was used \cite{SAeBL2014}. These two denominations are by far the most common in Germany and combined can be considered the rate of Christians.  Analysing all three distributions it becomes obvious that the determining factor is the share of Catholics in a given county. Regardless if only \textsc{hwc} or the sum of all studied objects (see Figure \ref{fig:germrel}) is used a threshold in the range between 30\,\% to 45\,\% is observed. Above this threshold a much higher rate of mapped nodes is recorded (note the logarithmic scale). While it is expected that targeted field research would lift the node densities in all regions it is assumed that the general trend would persist, as there are no indications of mapping biases related to the rate of Catholics within Germany.

\section{Bavarian Situation}
\begin{figure}
\centering
\includegraphics[width=1.00\columnwidth]{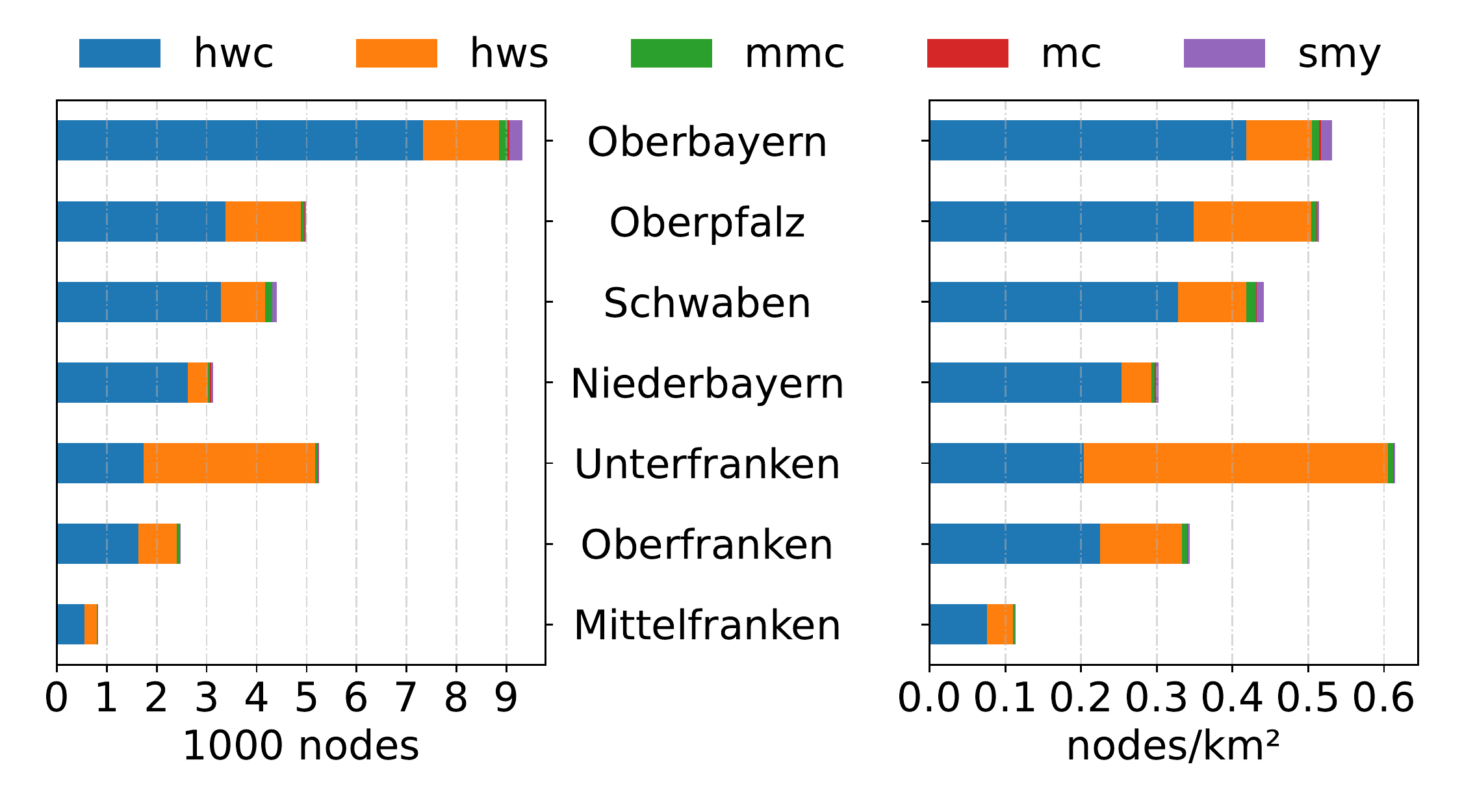}
\caption{Absolute count (left) and per area count (right) of \textsc{hwc}, \textsc{hws}, \textsc{mmc}, \textsc{mc}, \textsc{mtc}, and \textsc{smy} nodes for the seven Bavarian administrative districts, sorted by descending count of \textsc{hwc} nodes.}
\label{bavariannodes}
\end{figure}

When solely focusing on Bavaria, where about half of the German \textsc{hwc}-nodes (20393) are located, the North South trend remains (see Figure \ref{bavariannodes}). Additionally, as the Alps are mostly located in the South of Oberbayern (Upper Bavaria) and partly in Schwaben (Swabia), \textsc{smy} nodes are essentially all located here. Assuming there are no strong mapping biases towards certain tags, within Unterfranken a clear cultural difference is observed as this is the only administrative area with a higher count of \textsc{hws} than \textsc{hwc}. Confer the peak of \textsc{hws} around $50^\circ$\,N in Figure \ref{northsouthgermany}. 
	
Next, we will discuss the data quality and then see which analyses are possible with the available data. 

\subsection{Data quality}
Five measures to assess the quality of the data will be used.  First, the age of the node, i.e. the time since its last change with respect to May 23$^\mathrm{rd}$ 2023. Second, the version number of the node, i.e. the number of times it was changed. Third, the number of different contributing OSM-accounts in a given administrative district, which we will evaluate by calculating the respective Gini-coefficient \cite{Gini1912}. Fourth, we assess the average number of tags present. And last, we estimate the completeness by comparing with an official data set. 

Only \textsc{hwc}-nodes will be used for the quality assessment.

\subsubsection{Data age}
\begin{figure}
	\centering
		\includegraphics[width=1.00\columnwidth]{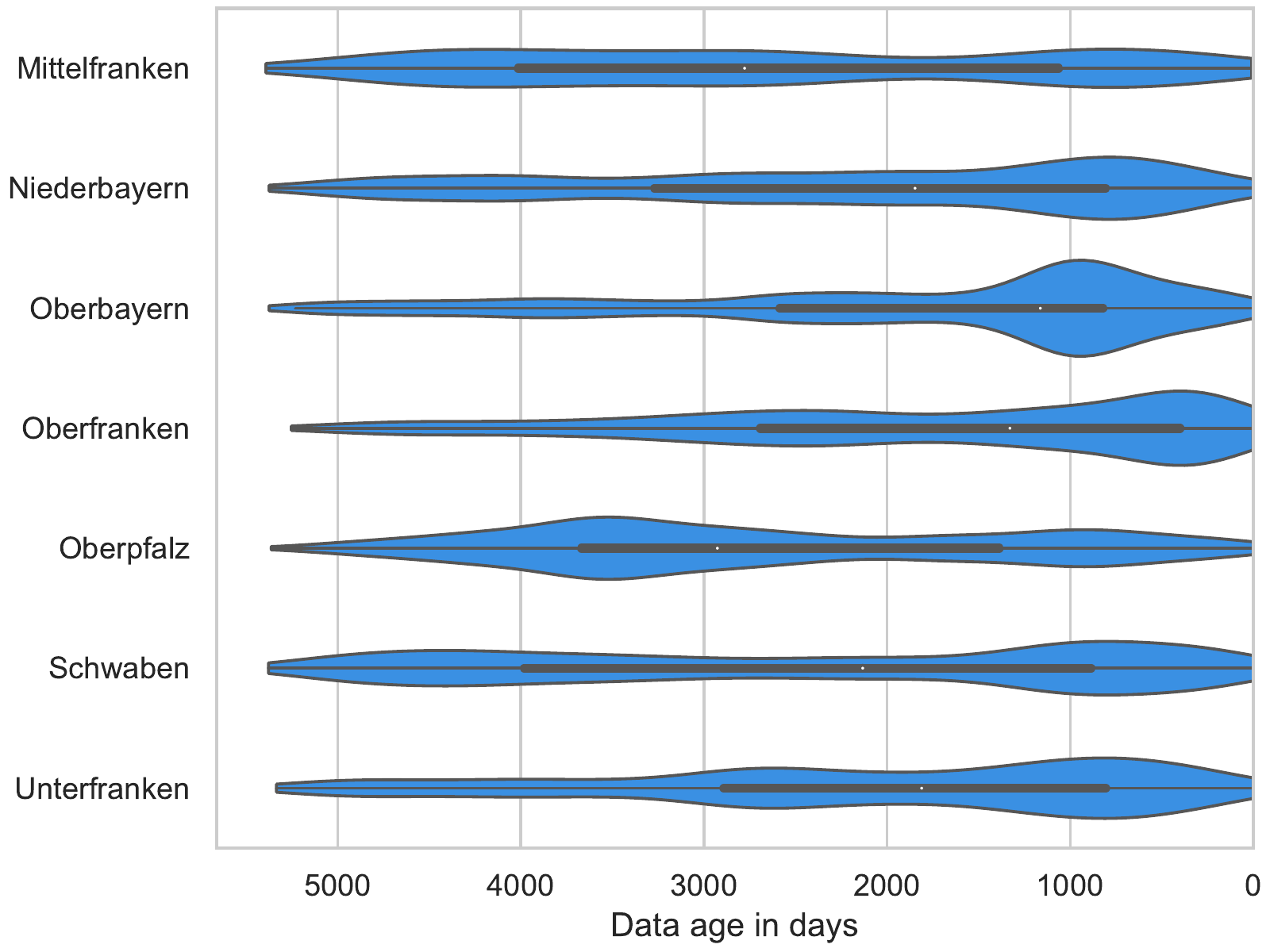}
	\caption{Data age distribution of \textsc{hwc} nodes in Bavaria by administrative district. The white dot denotes the media age, the box extends from the first quartile to the third quartile of the data. The whiskers extend no more than 1.5 times the interquartile range, ending at the farthest data point within that interval. The width of the violins indicate the relative number of entries.}
	\label{fig:dataage}
\end{figure}
For each of the seven administrative districts the distribution of the times since a node was last changed (confer Figure \ref{fig:dataage}) is determined. Everywhere about 14.5 years ago, or five year after the start \cite{osm_wiki_history} of the OpenStreetMap-project, the first node was created.  For comparison, the first post boxes appear at least four years earlier. This indicates that, as expected, the relative importance of wayside crosses is much lower than post boxes to the OpenStreetMap community. A wider study, including more tags would yield a relative importance of different aspects of culture and could also give interesting insights into differences among regions. The distribution is remarkably flat with several waves of addition over the last ten years. Also, in all administrative districts new data is added indicating, that the data in all districts is still incomplete. Over all, Mittelfranken (Middle Franconia) and the Oberpfalz (Upper Palatinate) have the oldest dataset with a median close to eight years. In contrast, Oberbayern has the newest data set with a median age of about 3 years.  As  the tag \textit{amenity}=\textit{post\_box} shows the same pattern it is also an indication that the contributing OSM community is more vital in Oberbayern than in the Oberpfalz. 
\begin{figure}
	\centering
		\includegraphics[width=1.00\columnwidth]{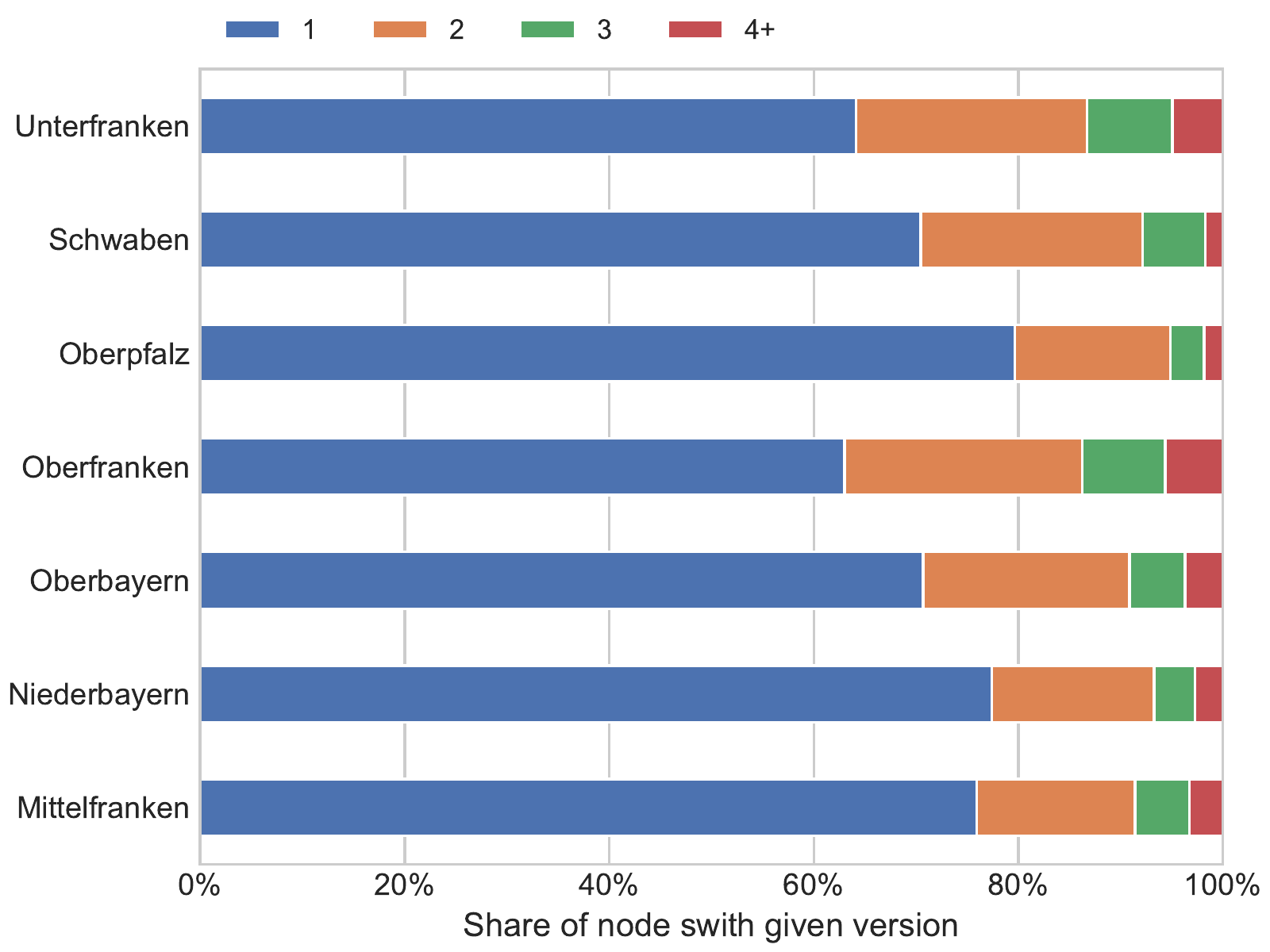}
	\caption{Percentage of \textsc{hwc} nodes with a given version number by administrative district in Bavaria. Version numbers $\geq 4$ are summed in the the $4+$ category. }
	\label{fig:admin_by_version}
\end{figure}

\subsubsection{Version count}
Figure \ref{fig:admin_by_version} summarizes the share of nodes with a given version by administrative area. Between 65\,\% and 80\,\% of the nodes where never changed since creation. Another 15\,\% to 20\,\% where changed twice, and 5\,\% to 10\,\% trice, the remaining more frequently. Typically three factors are expected to contribute to the version count. First, the ease of determining the position. As wayside crosses are often placed at way-crossings this factor is assumed to contribute little and thus yield lower version numbers. The second factor is the amount of meta data which sensibly can be added to a node. As we will show most of the additional information is either easy to determine in the first creation or is maybe not determinable at all. Thus also here a low count is expected. Last, dynamic objects tend to have higher version numbers, e.g. post boxes with changing collecting times, have only 20\,\% to 50\,\% of the nodes with only one version in the different administrative districts of Bavaria. As well, this is typically not the case for \textsc{hwc}-nodes. However, the rather low frequency of updates, points to the fact that \textsc{hwc} nodes are not the focus of mapping of the Bavarian OpenStreetMap community either.

\subsubsection{Points of view}
Data contributed by a single user tend to reflect a specific point of view, while different editors are expected to represent different facets. I.e. in the case of crosses, user1 might focus on the religious aspects, while user2 has a background in monument conservation. Anyhow, more users do not ensure different points of view, but make them more likely. 

\begin{table}
\small\sf\centering
\caption{Number of different contributors per administrative district.} 
\label{contributors}
\begin{tabular}{lr}
\toprule
Admin. district & Unique contributors \\
\midrule
Oberbayern    &  608 \\
Schwaben      &  354 \\
Oberpfalz     &  257 \\
Niederbayern  &  249 \\
Unterfranken  &  232 \\
Oberfranken   &  144 \\
Mittelfranken &  117 \\
\bottomrule
\end{tabular}
\end{table}
Although, in each administrative area contributions by more than a hundred individual users are recorded (see Table \ref{contributors}), there are only a handful that contributed the majority of the data in each area. Thus, all Gini coefficients are between 93\,\% and 99\,\%, i.e. in the regime of extreme inequality. Anyhow, this seems also to be characteristic for the OpenStreetMap-dataset as a whole, as a similar range (91\,\%-97\,\%) is found for post boxes. Still, in both cases the highest inequality is found in Oberfranken (Upper Franconia), while in comparison the most diverse distribution is found in Oberbayern. This again hints to a more vital community in Oberbayern.

\subsubsection{Data richness}
\begin{figure}
	\centering
		\includegraphics[width=1.00\columnwidth]{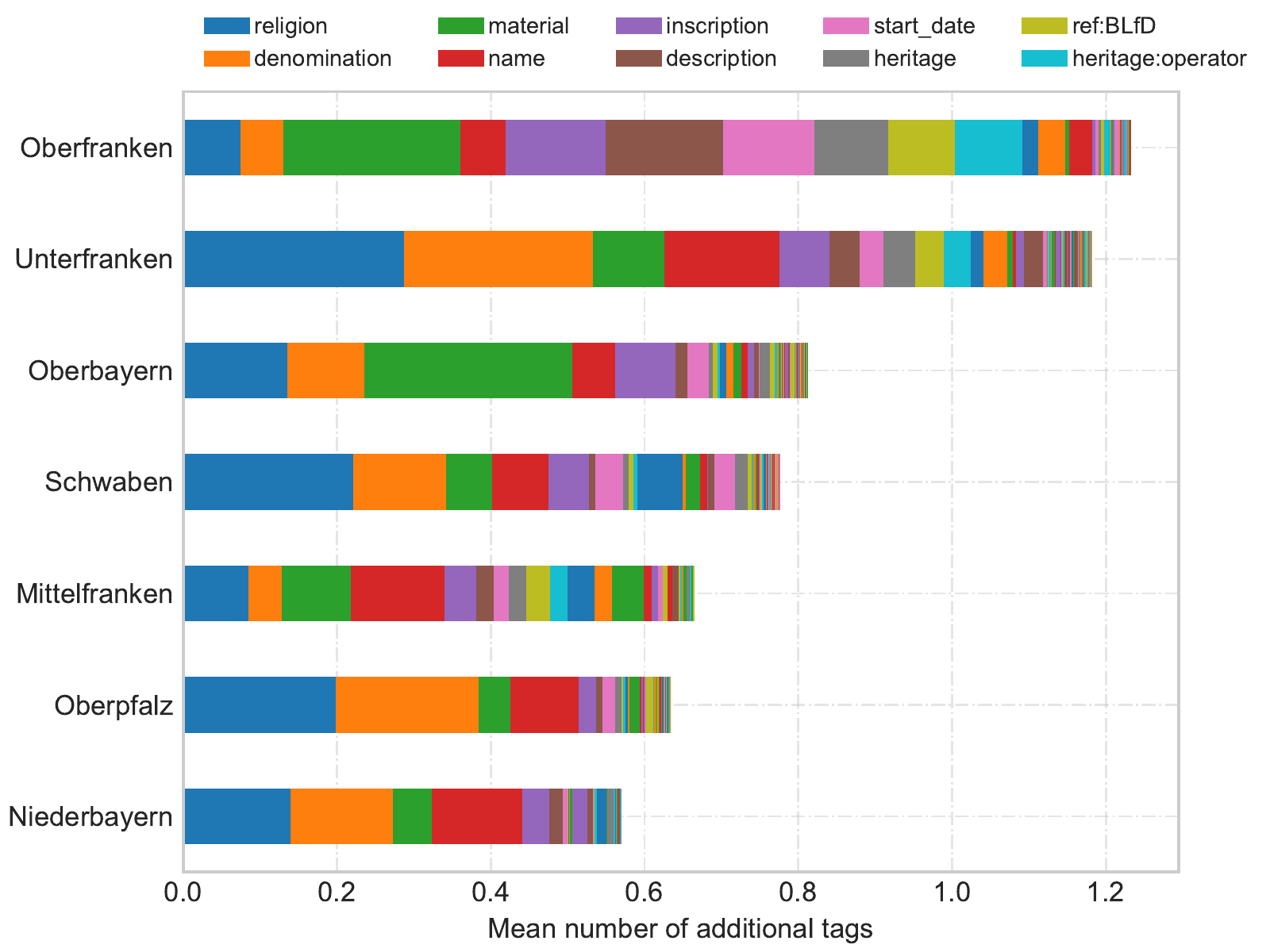}
	\caption{Mean number of additional descriptive tags for \textsc{hwc} nodes by administrative district in Bavaria. Each colour indicates a different tag but only the top 10 of Bavaria are explicitly named in the legend. }
	\label{fig:metastats}
\end{figure}

To asses the information content of the data, for each administrative district the mean number of additional descriptive tags is determined (see Figure \ref{fig:metastats}). These are all tags which are neither the main tag \textsc{hwc} nor one of the meta keys like \textsc{version} or \textsc{id}.

This analysis indicates, that the datasets of Ober- and Unterfranken (Upper and Lower Franconia) are on a relative scale the most feature rich with about 1.2 additional, descriptive tag on average. The most frequent key is \textsc{religion} which, e.g in Unterfranken, is recorded for about 25\% of the nodes. Here also for almost the same rate the \textsc{denomination}-key is present. However, the religion is not set for almost three quarters of the crosses. This most likely stems from two reasons: First, the contributor was not aware of the possibility of adding this information or regarded it as superfluous. The latter e.g. happens if the person entering the data assumes that crosses are by definition a sign of Christian religion. Second, the ground truth principle of OpenStreetMap, which prohibits entering information that is not verifiable on the ground. In this case the omission conveys, that it was not clear to the contributor if the cross had a religious or just a cultural meaning. This thought is very valid as was found e.g. in \cite{Przybylska2021}. While we expect the first reason to be the predominant reason for the \textsc{religion}-key, we expect a variant of the second reason to be causing the lower rate of the denomination tag; i.e. most contributors could not judge from the cross itself if it is e.g. \textsc{roman\_catholic} or \textsc{protestant}. Looking into the actual value we find that all crosses which exhibit a \textsc{religion}-key, are denoted as \textsc{christian}, and except for 14 nodes, all are tagged as \textsc{catholic} or \textsc{roman\_catholic} and one of the 14 was a misspelled \textsc{roman\_catholic}. This indicates a clear foundation of the rite within the roman catholic tradition and thus might explain the lower number of crosses in the protestant areas shown in Figure \ref{fig:germrel}. 
\label{datarichness}
  
Unfortunately, the \textsc{material}-key, denoting the main material of which the cross is made of, is only present to a larger extent in Oberbayern and Oberfranken. Thus, we can not differentiate the material distribution between the administrative districts in a sensible manner.

The fourth most used key is \textsc{name}, which denotes a popular or official name of the cross. We will discuss the names together with the \textsc{inscription}-key below in more detail. 

The key \textsc{start\_date}, contains the date when the cross was erected. It may contain accurate dates, to coarse estimates like “18th century”. While the data is sparse, and might be error prone as inscribed dates might not be the actual erection/scarification date, it still is indicative. With few exceptions, all dates are younger than 1850 and seem to form two eras. The first stretching from 1850 to the beginnings of the first world war, the second starting in the 1960s extending to today. Definitely more data would be needed to substantiate this. 

The tag cluster \textsc{heritage}, \textsc{ref:BLfD}, and \textsc{heritage:operator} all refer to official listings in heritage registers. In Bavaria this is in all cases the Bayerische Landesamt für Denkmalpflege (BLfD). 

Last in the top-10-key is \textsc{desciption}, indicating a further description of the object. These might be used by the community to enhance data richness.

\begin{figure}
	\centering
		\includegraphics[width=1.00\columnwidth]{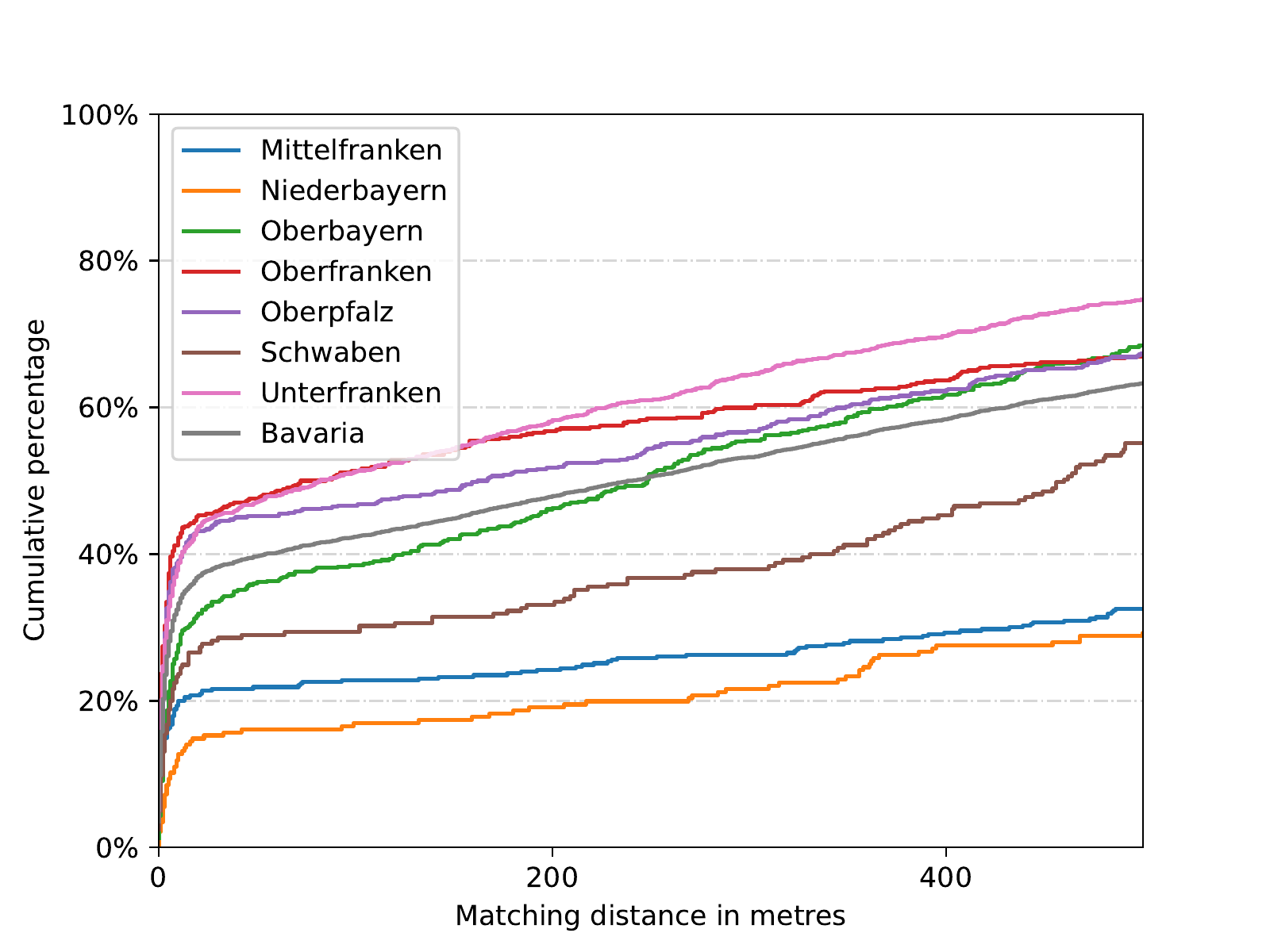}
	\caption{Percentage of matchable objects within a given search radius by administrative district in Bavaria}
	\label{fig:210511_blfd_}
\end{figure}

\paragraph{Comparison with official dataset}
While all prior indicators relied solely on the data itself, the last indicator uses an external, official dataset to measure the completeness: a list of the BLfD of the crosses registered as heritage in Bavaria \cite{blfdlist}. Even though, most crosses don’t qualify as heritage, many crosses are listed. Also it is  expected that listed crosses are in some aspects remarkable and thus might be recorded with a higher frequency in the OpenStreetMap-dataset. Thus, it is assumed that using listed crosses will yield a conservative and maybe slightly biased estimate of the data completeness. In the BLfD-dataset there are 4025 cross like objects, out of which 4022 carry a location, and of which 3459 are clearly some kind of cross, another 516 are not clearly marked, and another 50 would most likely be identified as something else, as the cross is only part of a church for example. Furthermore, in the BLfD-dataset all objects are recorded as small extended geometries. For the matching, for each of the 4022 geometries, the centroid was determined and the closest \textsc{hwc}, \textsc{hws}, \textsc{mmc}, \textsc{mc}, or \textsc{smy} node in the database was determined. In a second step the smallest distance is selected as the matched object. The result is summarized in Figure \ref{fig:210511_blfd_} up to a distance of 500\,m. Nodes farther away than this are considered not matched.

For Bavaria about 40\,\% of the matchable objects are matched within a couple of meters, for larger matching-radii an almost linear increase is found. Using the same matching radii as \cite{Balducci2021} of 30\,m, 50\,m, and 150\,m for comparability (see Table \ref{tab:matchcompare}) shows similar matching efficiencies on the 30\,m scale, but significant lower efficiencies on larger scales. This is expected as on the one side the position of crosses is typically well defined as discussed before and on the other hand museums are by far more extended objects. In contrary this also indicates that the completeness of the Bavarian cross data set in OpenStreetMap is by far less complete than the set of Italian museums. Still, about a third of the crosses, which at the same time also is less than 10\,\% of the complete data set also indicates that the dataset can be considered representative. 
\begin{table}
\small\sf\centering
\caption{Comparison of the matching efficiencies for cross objects in Bavaria with \cite{Balducci2021} for museums in Italy by matching distance.}
\label{tab:matchcompare}
\begin{tabular}{lrr}
\toprule
Matching radius &	Italian museums  &	BLfD crosses\\
\midrule
$ \mathrm{r} \leq\,30\,m$							&31.8\,\%	&38.2\,\%\\
$>30\,m \mathrm{r} \leq\,50\,m$		&14.8\,\%	&1.4\,\%\\
$>50\,m \mathrm{r} \leq\,150\,m$	&31.1\,\%	&5.2\,\%\\
$150\,m\leq \mathrm{r} $							&22.2\,\%	&55.2\,\%\\
\bottomrule
\end{tabular}
\end{table}

Also, for the individual administrative districts most objects are matched within a few metres. Still, the matching efficiencies differ significantly. While in the Oberfranken within 30\,metres 45.6\,\% can be matched, in Niederbayern (Lower Bavaria) this rate is only 15.2\,\%. Comparing the results with the additional tags discussed in the previous section, it is striking that Unterfranken and Oberfranken show a high rate of the \textsc{heritage}, \textsc{ref:BLfD}, and the \textsc{heritage:operator} tag, this is not true for the Oberpfalz -- again indicating a lower data quality in this administrative district. 

So in summary all data quality indicators suggest that the data in the Oberpfalz and Niedrbayern have the lowest confidence, while that of Oberbayern and Unterfranken are the ones with the best foundation. 

\subsection{Analyses}
Based on our basic understanding of the data quality we will now first derive an absolute count of crosses for Bavaria. Second, we will discuss the linguistic characteristics of the \textsc{name} and \textsc{inscription} key values. Third, we discuss the used materials.

\subsubsection{Absolute count}
\begin{figure}
	\centering
		\includegraphics[width=1.00\columnwidth]{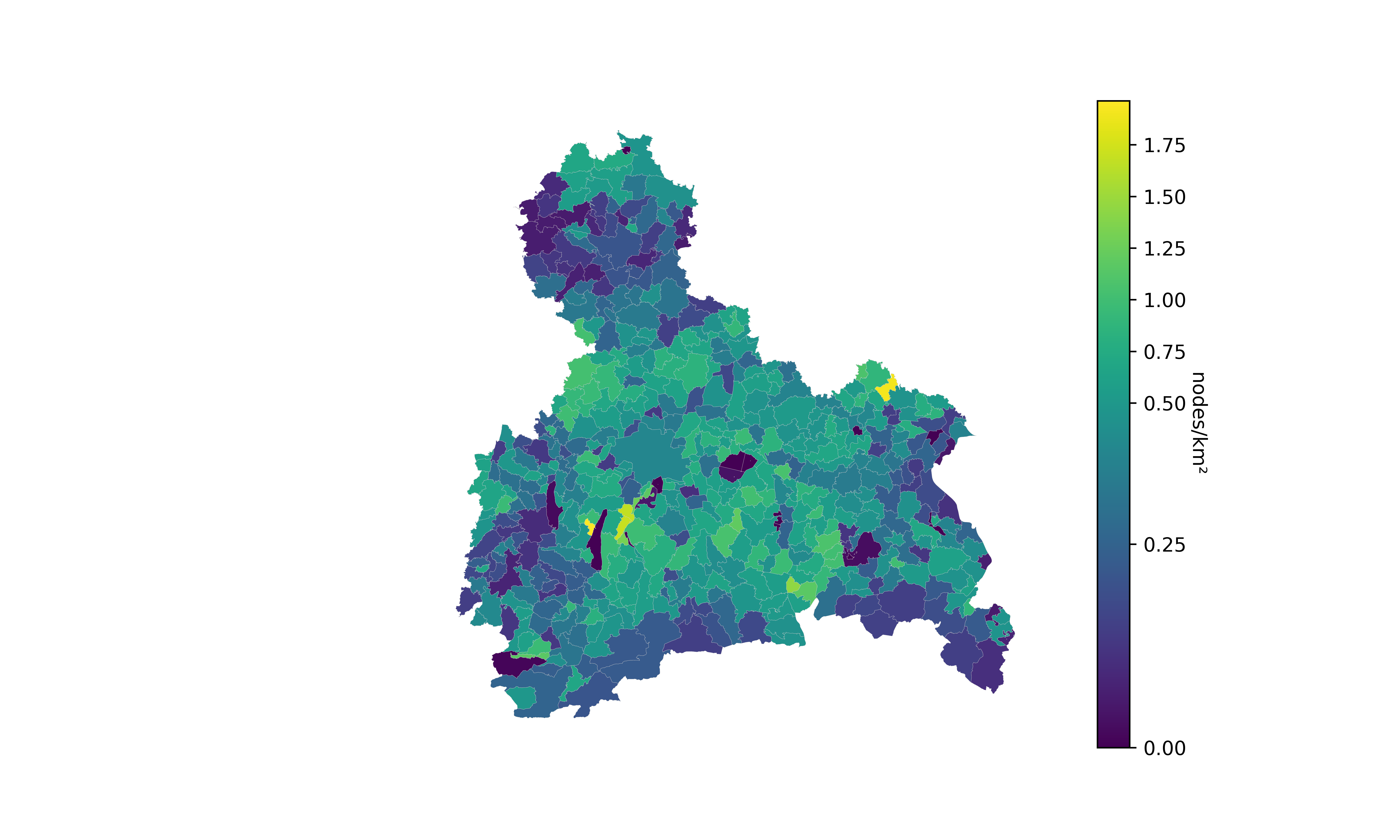}
	\caption{Mapped \textsc{hwc} nodes per km$^2$ per county in Oberbayern.}
	\label{fig:Oberbayern_map}
\end{figure}
For Oberbayern, the author tried to estimate the completeness of the data by conducting field research by car, by bike and by foot. It was found that, with the exception of the county Emmerting, it was easy to add several crosses in every county. On the other end, it was found, that finding more than about 0.3 crosses/km$^2$ was only possible with extensive research in most counties. More extensive research yielded densities of at least 0.8 crosses/km$^2$. Acribic research, which was conducted in Schäftlarn, Baierbrunn and Icking, yielded up to 1.7 crosses/km$^2$ (see also Figure \ref{fig:Oberbayern_map}). Averaging all counties in Oberbayern a value of $(0.5\pm 0.3)$ crosses/km$^2$ is found. As it is assumed from the observations before, that data is incomplete the actual average is assumed to be above 0.8 but below 1.8 crosses/km$^2$. Thus it is estimated that there are 14000 to 32000 crosses in Oberbayern and that about one third to one sixth of them are mapped. 

As no field research was conducted in the other administrative districts the findings will be extrapolated for the other administrative districts using the matching efficiencies found for the BLfD-dataset. Here we will employ the efficiency at 50\,m as upper bound and the one at 150\,m as lower bound, yielding between 60000 and 75000 crosses in Bavaria out of which one fourth to about one third is mapped. As another proxy for comparison, the number of farms in Bavaria was about 83100 in 2019 \cite{BayerischesStaatsministeriumfuerErnaehrung2020}. Historically many farms in Bavaria had a cross and during the field trips this was found to be still true especially for smaller farms. Given the size structure, it is estimated that about one fourth to one third of the farms should have a cross. This would give a lower limit of about 20000 to 25000 crosses. As besides farms there are many other crosses, the number of 60000 to 75000 crosses estimated above seems to be a good estimate.

\subsection{Analysis of the name and inscriptions}
Two data fields, the \textsc{name} and the \textsc{inscription}-key, contain text. The former an official name of the cross and the latter anything, which is written on the cross itself. All data contained in these two keys of \textsc{hwc} nodes in Bavaria was classified using nltk \cite{Bird2009} and HanTa \cite{Wartena2019} to identify words, lemmas and parts of speech (PoS). In eleven cases the inscriptions were longer than the maximum field length of 255 or were recorded with additional detail, e.\,g. facing south.  In the first cases several mitigation strategies are used. One is to introduce new tags like \textsc{inscription\_1}. Still, for this analysis these extensions were discarded and only the 1431 standard inscriptions were used.
\begin{figure}
	\centering
		\includegraphics[width=1.00\columnwidth]{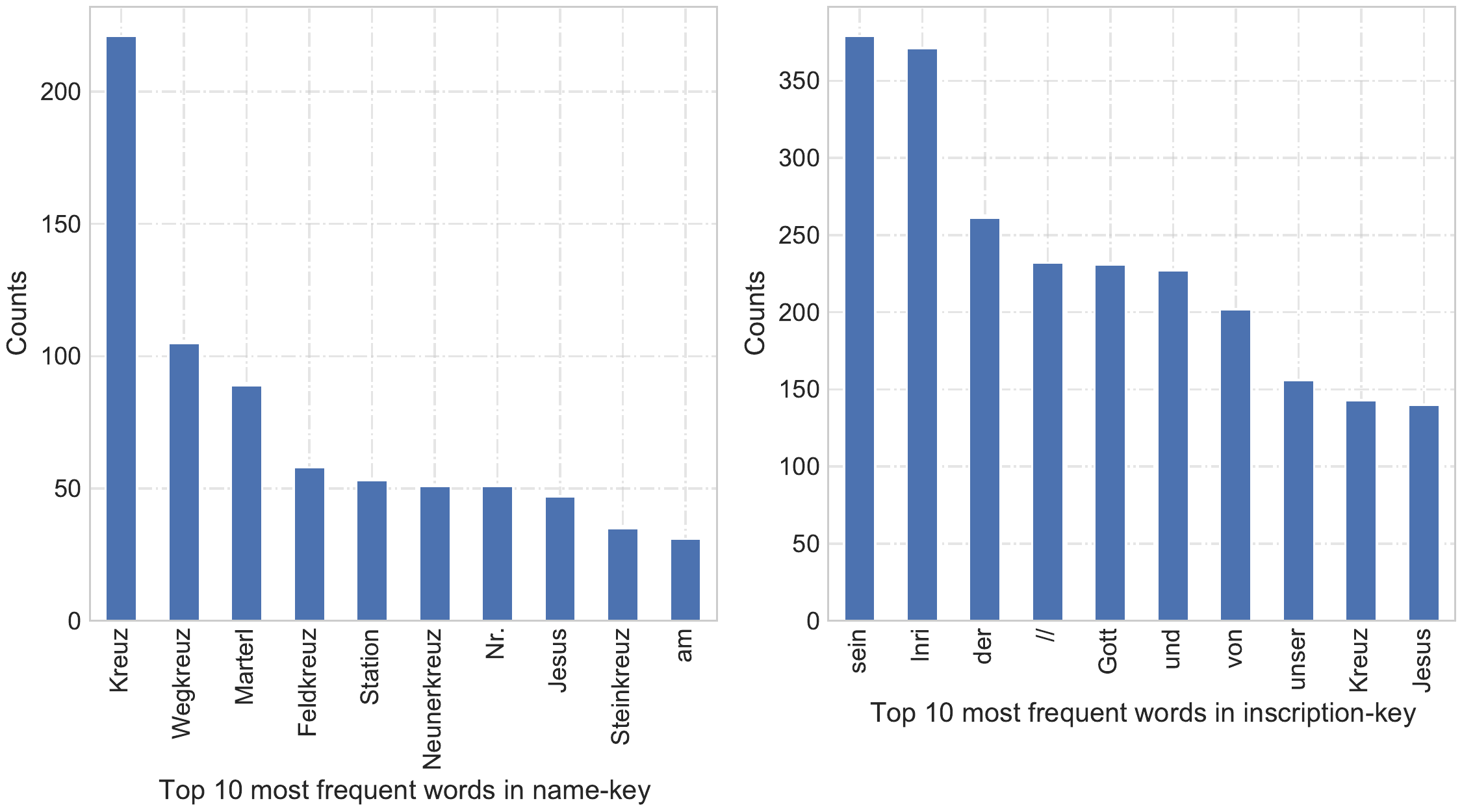}
	\caption{Top 10 most frequent lemmas, omitting punctuation, in the \textsc{name} (left) and \textsc{inscription} (right) key-value}
	\label{fig:lemma}
\end{figure}

Out of the 1555 names recorded, by definition many are unique referring to persons or events connected to the cross. Anyhow, lot of the data contained in the \textsc{name} field (see Figure \ref{fig:lemma} (left)) is just a German repetition of the tag \textsc{hwc}, i.e \textsl{Kreuz, Wegkreuz, Marterl, Feldkreuz,} or \textsl{Steinkreuz}. While Steinkreuz (stone cross) and the less frequent Holzkreuz (wooden cross) might be considered carrying some additional information, which should be listed in the material key, most of these keys likely should simply be deleted as they are not the actual \textsc{name} of the cross. The value \textsl{Jesus} seems to hint to a Jesus affixed to the cross. Here the cross can be considered Christian. Anyhow, the corresponding \textsc{religion}-key is very sparsely found. This hints to potential for improvement of the data. Another name denotes that the object is part of a way of crosses (\textsl{Station}). Leaving only one entry in the top-10, which contains actual information: \textsl{Neunerkreuz}\footnote{\textsl{am} is just a German preposition} These are crosses erected in commemoration of the battle on the $24^{th}$ April 1809 in Neumarkt-St. Veit, which are also called \textsl{9er-Kreuz \cite{Gruber2019}}. As expected, all of these crosses are located near Neumarkt-St. Veit. They also account for all the \textsl{Nr.} (number) entries. 

\begin{figure}
	\centering
		\includegraphics[width=1.00\columnwidth]{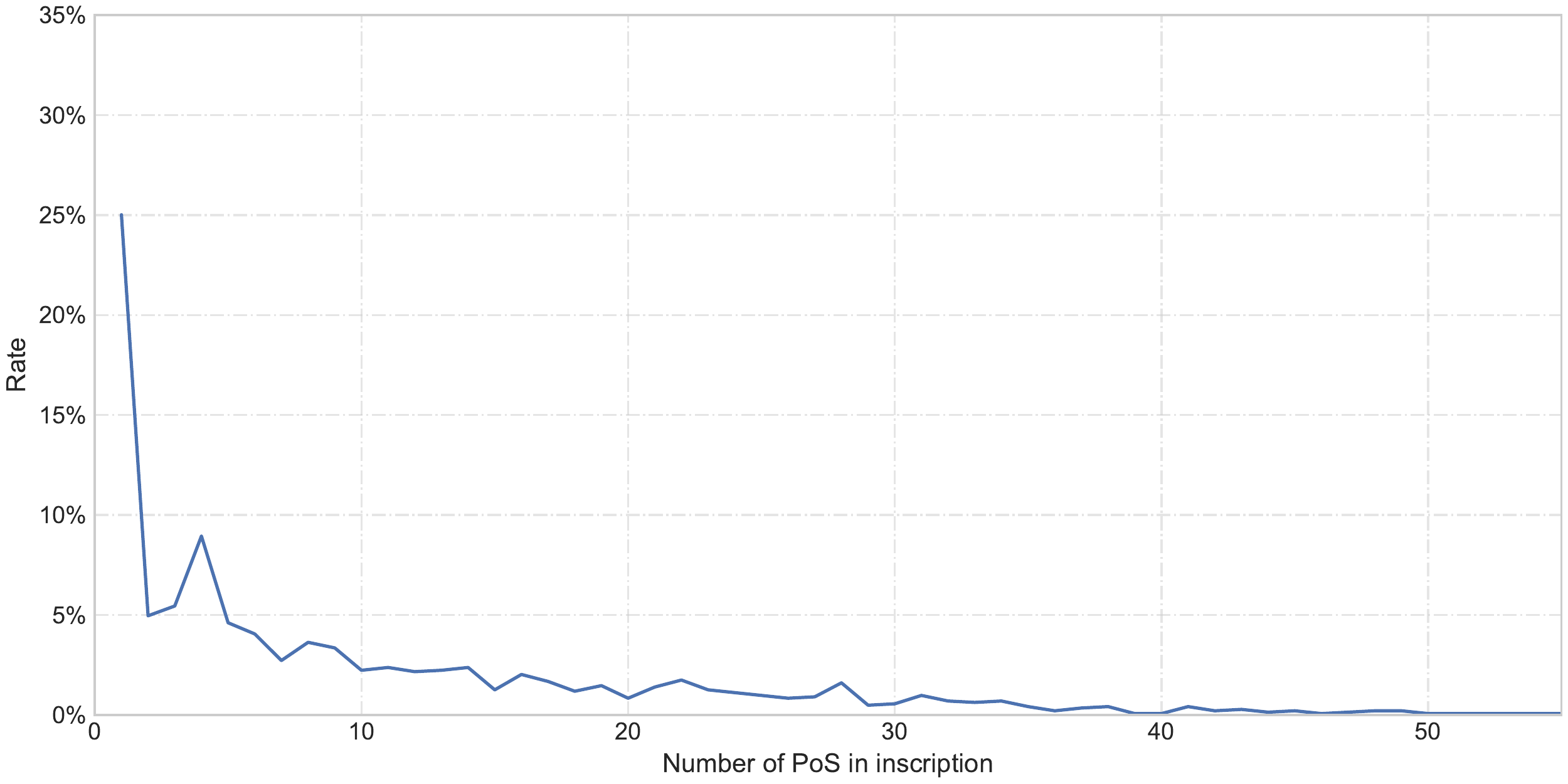}
	\caption{Share of values of given length in the values of \textsc{inscription}-keys}
	\label{fig:leninscription}
\end{figure}
Two distinct classes characterize the \textsc{inscription} key (see Figure \ref{fig:leninscription}). In the larger class the crosses carry only the four letter acronym INRI (Iesus Nazarenus Rex Iudaeorum), some add a year and/or name in addition. The second class bare a longer text. This often is a poem or proverb praising God, Jesus, Marry, or another Saint but sometimes also a text describing the situation why the cross was erected. Still, the texts are very diverse which is reflected by the fact that besides God and Jesus none of the deities praised are found within the top-10 - even Jesus is only on rank 10. Still at least two words, "unser" (our) and "hier" (here), indicate that the text on the cross speaks about something related to the erecting entity or the place where the cross was erected. 

\begin{figure}
	\centering
		\includegraphics[width=1.00\columnwidth]{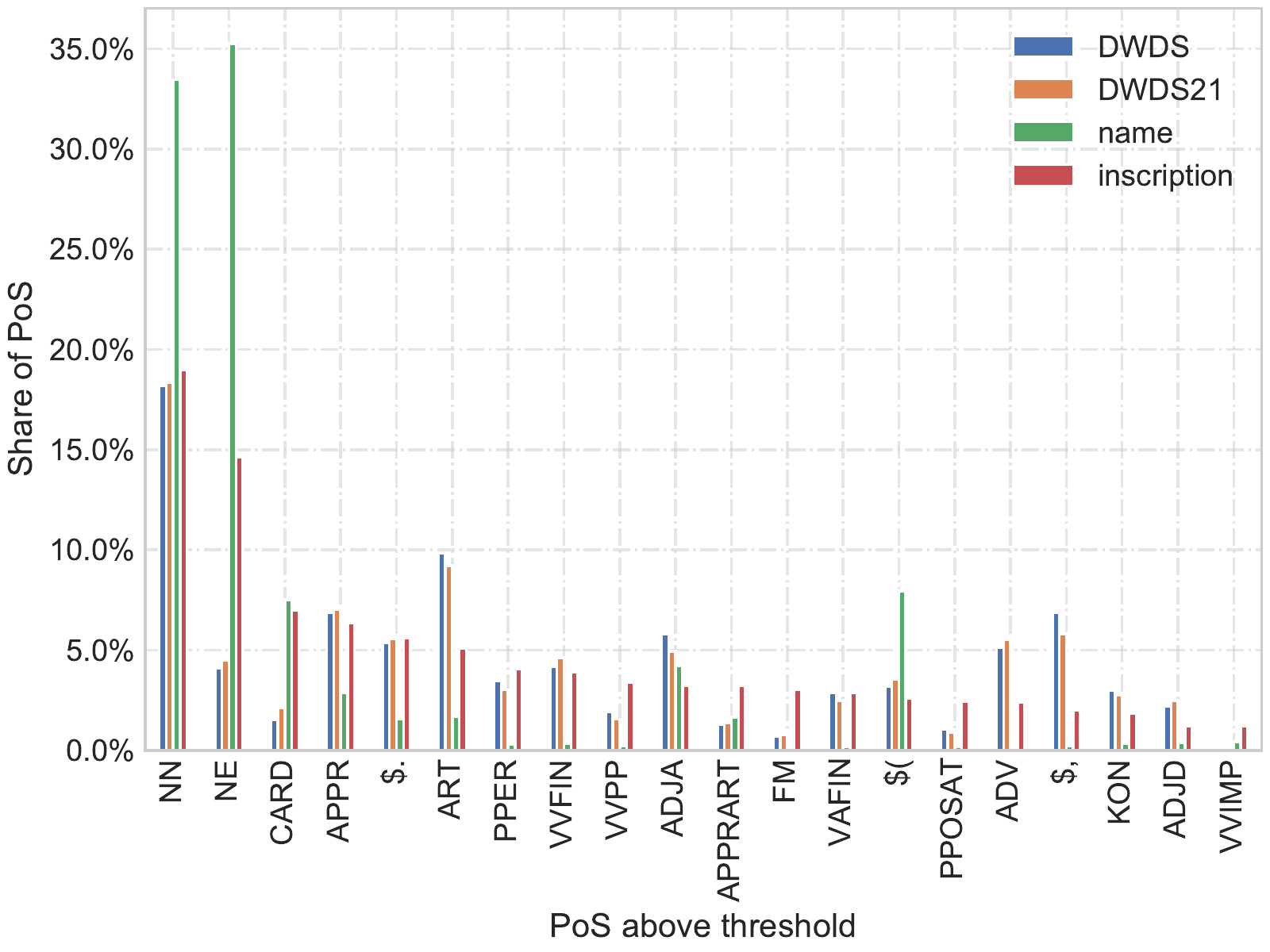}
	\caption{Relative PoS distribution with respect to the corresponidng corpus for all PoS which acount for at least 1\,\% of the PoS within the \textsc{name} or \textsc{inscription} key. Study area is Bavaria.}
	\label{fig:posstats_total_distribution}
\end{figure}
To assess the content quantitatively the distribution of the most frequent PoS is compared with respect to the distribution found in two generic German corpora compiled by the Digitales Wörterbuch der deutschen Sprache (DWDS) \cite{Klein2010}. The first corpus spans the time period 1900 to 1999 \cite{Geyken2007}, the second (DWDS21), the time period 2000 to 2010. In total, adding up all absolute differences between the PoS frequencies in the two different corpora and the name and inscription tag, it is found that the two corpora only differ by about 0.1. This difference will also be employed to measure the uncertainty. With respect to the inscription tag, both reference corpora exhibit a difference of $0.5\pm 0.1$. For the name tag the difference is $1.1\pm 0.1$ for the DWDS corpus and $1.0\pm 0.1$ for the DWDS21 corpus.

The large difference for the \textsc{name} key values is obvious as it mainly contains single words or fragments of sentences and thus is dominated by nouns, normal nouns (NN) and proper nouns (NE), while the other categories are very distinctively populated. The second most prominent PoS are inter-punctuations (\$,, \$(, \$.) and cardinal numbers (CARD). The last two PoS found in notable number in the \textsc{name} key values are adjectives (ADJA and ADJD) as well as prepositions (APP*). In contrast, the \textsc{inscription}-distribution is much closer to a typical distribution for German texts -- $0.5\pm 0.1$ for the DWDS and DWDS21corpus. Still there are notable differences. First, as many crosses where erected in commemoration of specific persons and the name INRI is very dominant, the rate of NE is about three times higher than in standard corpora. In addition, the number of cardinals is significantly higher as expected. And while articles (ART) are less frequent articles combined with prepositions (APPRART) are more frequent. 

\subsection{Used Building Materials}
The \textsc{material}-key “describes the main material of a physical feature” \cite{osm_wiki_material}. In total 18 different materials are used to describe the \textsc{hwc} in Bavaria. They can be clustered in three main groups: wood, stone, and metal.  By far the majority of the crosses with recorded material are wooden, with no distinction between different wood types, as obviously the contributors are not able to distinguish weathered wood types. Among the stones, especially sandstone is called out explicitly. Metal crosses with an explicit tagging are mostly different forms of iron. Another 49 "materials" are present which are either describing combinations of materials or are simply german terms. While, the former should be solved by a better tagging sheme, the latter is incorrect acording to the OSM-rules.
\begin{figure}
	\centering
		\includegraphics[width=1.00\columnwidth]{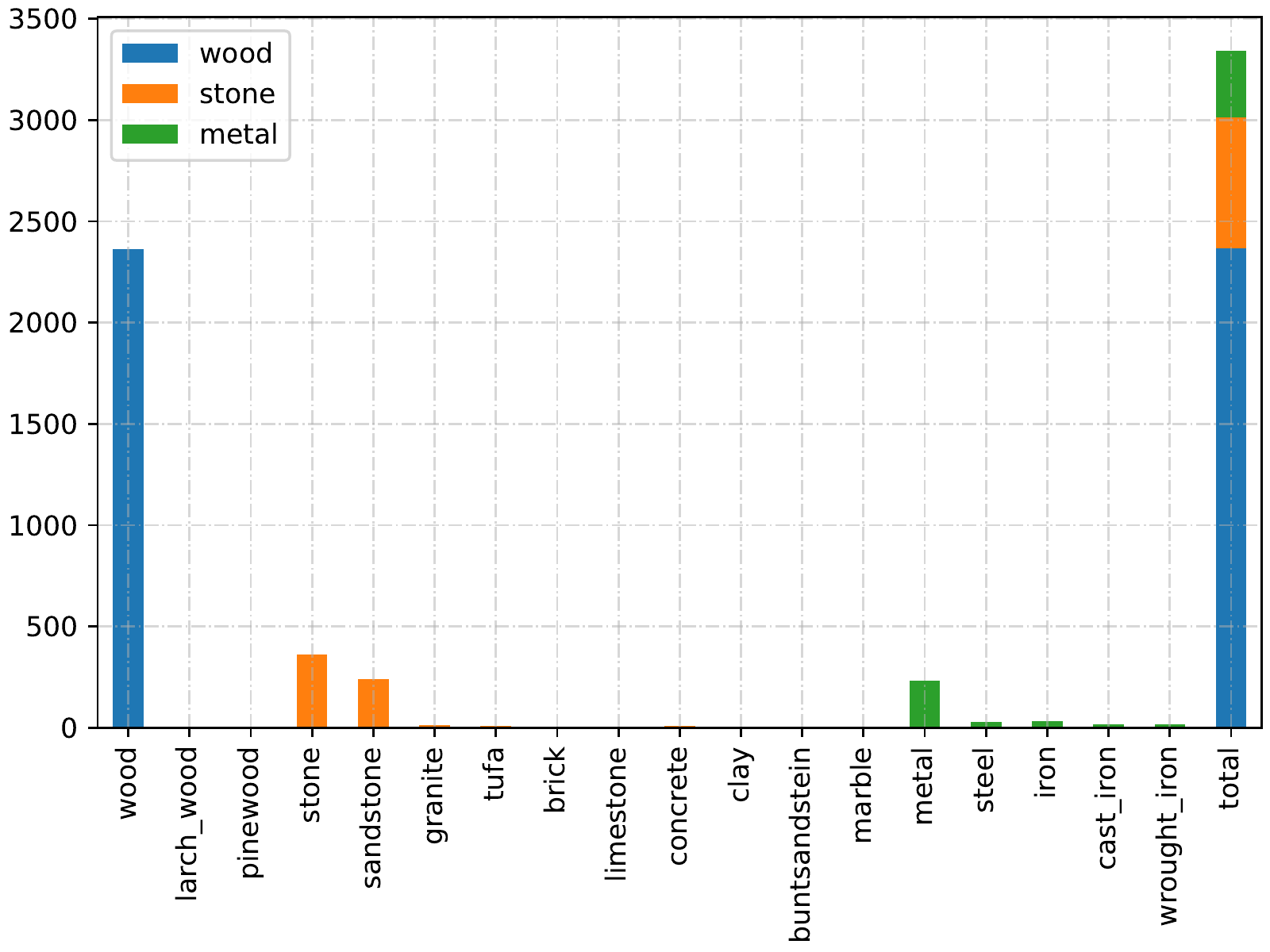}
	\caption{Count different values in the \textsc{material}-key for  \textsc{hwc} nodes in Bavaria}
	\label{fig:material}
\end{figure}

Although, only about 17\,\% of the crosses are tagged with a material, it is assumed that the distribution is representative. The main bias is probably, that materials that are perceived uncommon would be registered more frequently. Both, stone and metal account for 10-20\% each. A strong bias effect would result in an lower number in reality. 

Further analyses of the material would need a better distributed data. In addition, due to the low statistics and reliability of the \texttt{start\_date} key, it is not possible to infer changes of material over time reliably, as they are for example observed in Quebec, Canada \cite{Kaell2015}.
  
\section{Conclusion}
When assessing data quality in the terminology of the ISO norms for geographic data quality ISO 19113 \cite{ISO19113} and ISO 19157 \cite {ISO19157} as documented in \cite{Antoniou_2015} we conclude:
Exploiting the vast amounts of data found in the OpenStreetMap database for studies outside of cartographic mainstream still is not universally possible without due care. This has several reasons: First, the \textbf{completeness} of the data varies considerably between different countries and even within countries. In addition, also the richness of the data varies considerably. Third, due to the high concentration on certain contributors, the data might be biased considerably. Fourth, especially for non-mainstream \textsc{key=value} combinations the folksonomy is sometimes not well defined yielding challenges for the \textbf{logical consistency} and the \textbf{thematic accuracy}.

Still, many of these challenges can be overcome if for certain regions or subsets of the data reference points can be obtained, e.g. by field research, a reference dataset or other sources. Also the biases might be irrelevant, if the research goals are concisely chosen. For example a study on the distribution of diaper changing tables in restaurants will most likely be problematic, given the social composition of the OpenStreetMap-community, but a question on the average age of street\_lamps would most likely be unaffected by social biases if the data is present. Furthermore, the question of the folksonomy can be solved by understanding the different keys and tags used by using the standard OpenStreetMap-tools, like the wiki.

Furthermore the \textbf{positional accuracy} was found to be in general very good and also the \textbf{temporal accuracy} can be considered good, as the data is constantly updated. Taking these findings together the \textbf{usability} of the data is mostly only given, if a good understanding of the mechanism of its creation can be gained. 

\section*{Acknowledgements}
Map data copyrighted OpenStreetMap contributors and available from https://www.openstreetmap.org. The Shape-file for the map of bavaria in Figure \ref{fig:Oberbayern_map} is revision 01.08.2018 and copyrighted by Bayerische Vermessungsverwaltung – Verwaltungsgebiete under cc-by. The BLfD-list of crosses  is in the revision of 31.03.2021 and is copyrighted by the BLfD.

\printbibliography

\end{document}